# Design of High-Q Passband Filters Implemented through Multipolar All-Dielectric Metasurfaces

Alessio Monti, *Senior Member, IEEE*, Andrea Alù, *Fellow, IEEE*, Alessandro Toscano, *Senior Member, IEEE* and Filiberto Bilotti, *Fellow, IEEE*

*Abstract*— **We propose a novel class of ultra-thin high-Q passband filters designed by properly combining different multipolar resonances sustained by an all-dielectric metasurface. A rigorous analytical model, based on surface impedance homogenization and accounting for the effects of both dipolar and the quadrupolar contributions to the overall scattering response, is derived and verified through numerical simulations. Then, it is described how it is possible to engineer the interactions between dipoles and quadrupoles in a metasurface made by core-shell spherical elements to design ultrathin and broadband dielectric mirrors with a narrow transmission band. The proposed filters exhibit high Q-factor resonances and can be implemented using realistic materials at either microwave or optical frequencies. Finally, we discuss how the proposed dielectric filters can be used to design self-filtering aperture antennas exhibiting higher out-of-band selectivity compared to those implemented through metallic resonators.**

## I. Introduction

Differently from conventional microwave filters that act on the guided modes of a waveguide, the design of filters for free propagating electromagnetic (EM) waves, which are commonly referred to as *space filters*, is usually more complex, due to the need to control the response for many impinging angles and for differently polarized fields. Arguably, the first examples of space filters are based on the so-called *frequency selective surfaces*. These structures, formed by a periodic arrangement of metal patches, exhibit a rapid frequency variation of the reflection and transmission coefficients around their resonance and can be designed to behave as pass-band or notch filters for EM waves [1]. Many designs, based on multilayered geometries for more complex frequency responses, have been proposed over the years [2]-[5]. More recently, it has been shown [6] that another class of periodic surfaces, *i.e.*, the metasurfaces, are also able to control transmission and reflection independently and, thus, exhibit a frequency-selective behavior [7]. Differently from FSS, metasurfaces are characterized by subwavelength periodicities and by a lower sensitivity with respect to the impinging angle.

A particular class of metasurfaces, which is catching the interest of different scientific communities in the last years, is known as *all-dielectric metasurfaces* [8]. These structures are composed by high-index dielectric meta-atoms of various shapes supporting both electric and magnetic resonances.

Manuscript received July 27, 2020. This work has been developed in the framework of the activities of the research contract MANTLES, funded by the Italian Ministry of Education, University and Research as a PRIN 2017 project (protocol number 2017BHFZKH), and by the Air Force Office of Scientific Research.

A. Monti is with the Niccolò Cusano University, Rome, Italy (corresponding author, e-mail: alessio.monti@unicusano.it).

A. Alù is with the Photonics Initiative, Advanced Science Research Center, City University of New York, 85 New York, NY 10031, USA, and with the Physics Program, Graduate Center, City University of New York, New York, NY 10016, USA.

F. Bilotti and A. Toscano are with the Department of Engineering of ROMA TRE University, 00146 Rome, Italy.

Color versions of one or more of the figures in this communication are available online at http://ieeexplore.ieee.org.

As they do not contain any metallic part, all-dielectric metasurfaces are well suited for high-frequency applications or when high-power EM fields are considered. In the last years, the possibility enabled by these structures have been widely investigated and it has been shown how the interaction between the electric and magnetic dipole allows several intriguing effects, including the design of non-metallic mirrors [8], invisibility cloaks [10] or devices for field transformation [11]. All-dielectric metasurfaces have been mainly proposed for nanophotonic applications, due to the unavailability of high-conductivity materials at infrared and optical frequencies. However, the ever-growing demand for the next-generation communication systems to operate in the millimeter and sub-THz frequency ranges has driven the interest in all-dielectric metasurfaces also at lower frequencies [12]-[15].

This work aims at exploring a novel approach to achieve filtering of EM waves based on the properties of all-dielectric metasurfaces. The idea is to exploit and engineer the complex multipolar response of properly designed dielectric meta-atoms in order to obtain a narrow passband behavior within a wide reflection bandwidth. As it will be shown, this effect is made possible by combining both dipolar and quadrupolar resonances sustained by core-shell particles. An analytical model, based on a surface impedance homogenization approach, is put forward, and its effectiveness is validated through full-wave simulations. In addition, we show how a proper synthesis of a core-shell particle allows designing self-filtering aperture antennas exhibiting a narrowband behavior, with important advantages in terms of signal-to-noise ratio and rejection of interfering signals. Compared to conventional dielectric resonator filters, designed to operate for guided waves [16], the proposed components are intended for free-space EM waves and are extremely thin, while ensuring excellent selectivity performances.

## II. Analytical Modelling

### A. EM behavior of the individual meta-atom

The geometry of the individual element we consider here is shown in the inset of Fig. 1 and it consists of a core-shell spherical particle. The outer shell has a radius $r_s$ and is made of a high-index dielectric with relative permittivity $\varepsilon_s$, whereas the inner core has radius $r_c$ and is made of a low-index dielectric with relative permittivity $\varepsilon_c$.

As well-known, regular high-index spherical particles with radius comparable to the dielectric half-wavelength exhibit a magnetic and an electric dipole resonance occurring at close frequencies [8]. As it can be appreciated in Fig. 1 (solid lines), where the amplitudes of the first three scattering coefficients ($a_1$, $b_1$ and $b_2$) of a spherical particle with $\varepsilon_s = \varepsilon_h \gg 1$ are shown, the resonances of the higher order multipoles occur above the dipolar electric resonance.

Moving from the uniform spheres to the core-shell geometry with a strong permittivity contrast allows engineering the spectral position of the electric resonance, while keeping unaffected the magnetic resonance. This is confirmed by the analysis of the scattering coefficients of a particle with $\varepsilon_c = 0.1\varepsilon_s$ shown in Fig. 1 (dashed lines) for different values of the core radius $r_c$. The physical reasons behind this behavior can be understood by considering the different flow of the displacement currents inside a dielectric sphere for the electric and magnetic dipolar resonances [17]. In the former case, in fact, the currents are concentrated in the central region of the sphere and any modification of its geometry or material properties strongly affects their resonance frequency. In the latter case, instead, the displacement currents flow in the outer region of the sphere and their distribution is not affected by the core-shell configuration.

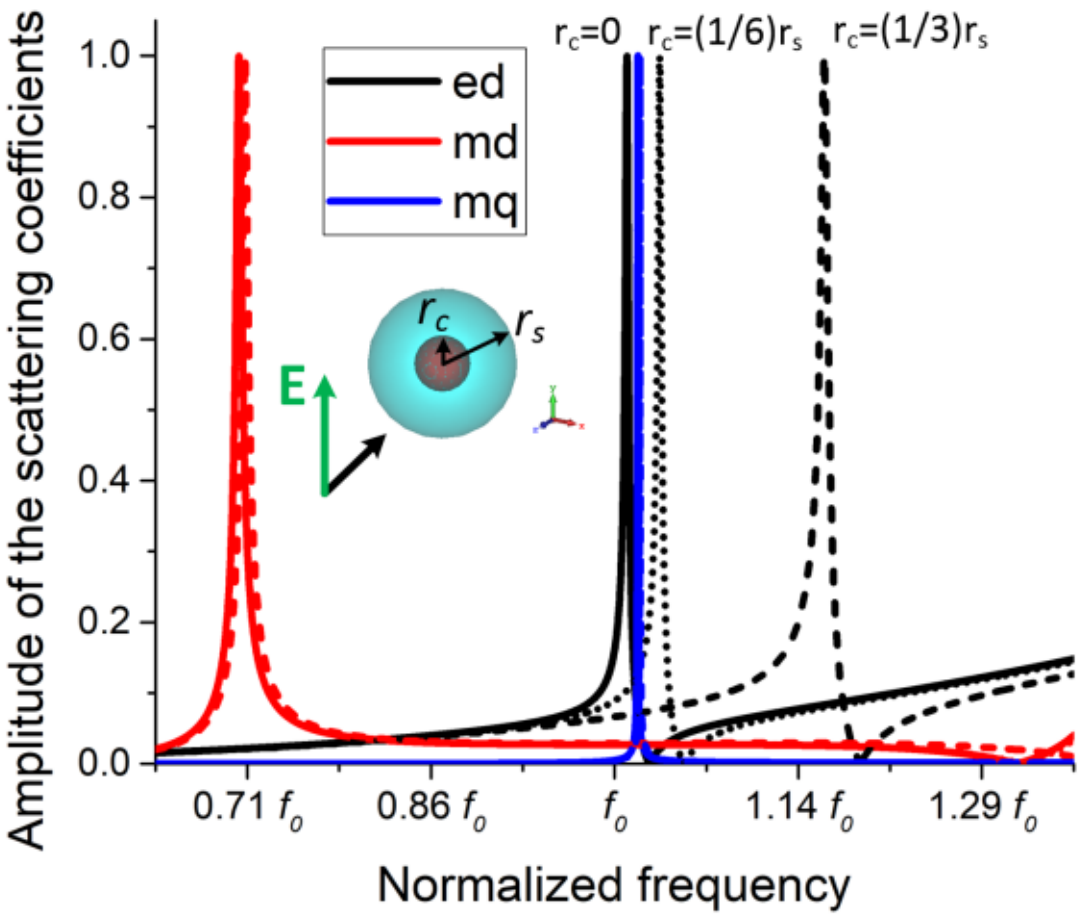

Fig. 1. Amplitude of the first three scattering coefficients of a core-shell spherical particle in different scenarios: (*i*) $r_c$ = 0 (solid lines); (*ii*) $r_c$ = (1/6)×$r_s$ (dashed lines); (*iii*) $r_c$ = (1/3)×$r_s$ (dot-dashed lines). In all cases, $r_s$ = 0.07$\lambda_0$, being $\lambda_0$ the resonance wavelength of the magnetic quadrupole, and $\varepsilon_c$ =0.1×$\varepsilon_s$. Red, black and blue colors refer to the magnetic dipole (*md*), electric dipole (*ed*) and magnetic quadrupole (*mq*).

As it can be appreciated in Fig. 2, very similar results can be also obtained by changing the permittivity of the spherical core once fixing its radius: as the core permittivity decreases, the resonance of the electric dipole is progressively shifted towards high frequencies, whereas the resonance frequencies of the magnetic-type modes are almost unaffected.

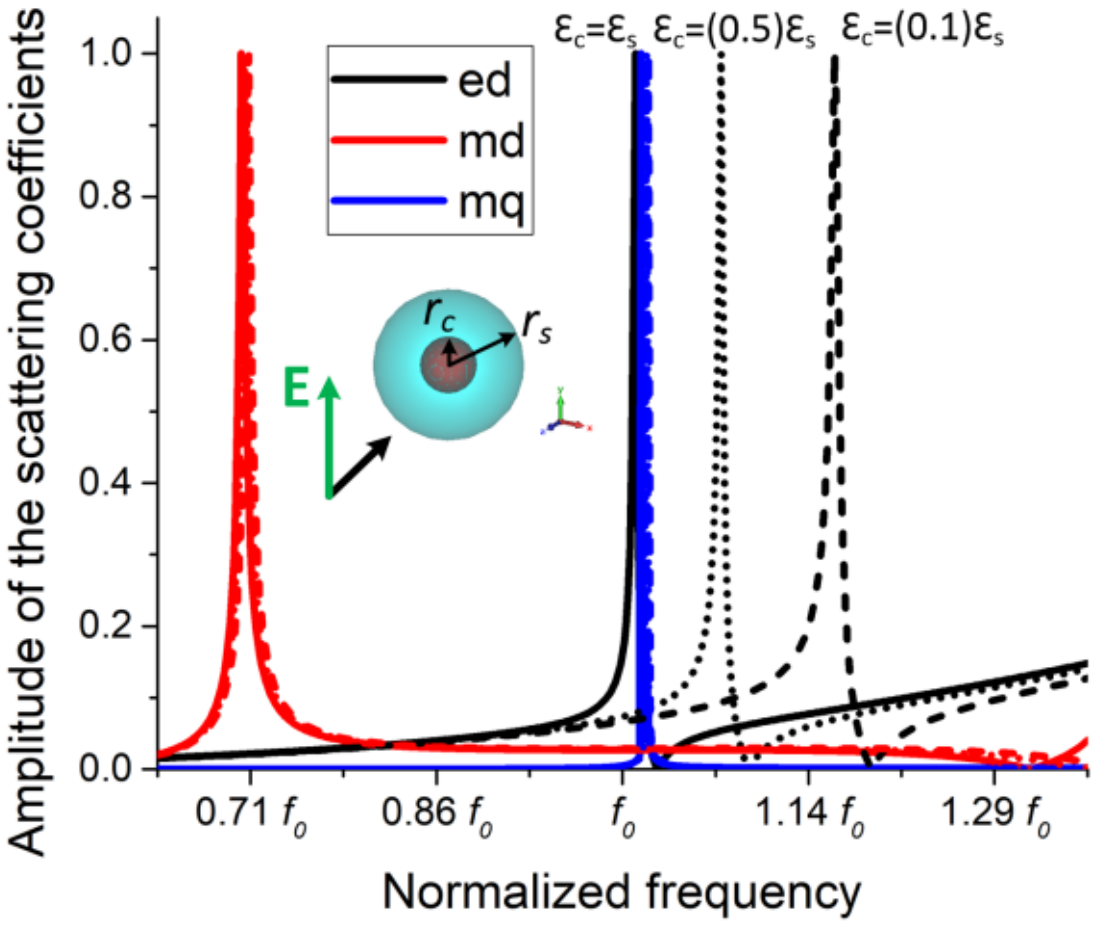

Fig. 2. Amplitude of the first three scattering coefficients of a core-shell spherical particle in different scenarios: (*i*) $\varepsilon_c = \varepsilon_s$ (solid lines); (*ii*) $\varepsilon_c$ = (0.5)× $\varepsilon_s$ (dashed lines); (*iii*) $\varepsilon_c$ = (0.1)×$\varepsilon_s$ (dot-dashed lines). In all the cases, $r_s$ = 0.07$\lambda_0$, being $\lambda_0$ the resonance wavelength of the magnetic quadrupole, and $r_c$ =(0.33)×$r_s$.

As it will be clearer later, the possibility of tuning only the resonance frequency of the electric dipole by acting on the permittivity or on the radius of the core is at the basis of the design of passband filters. In the filter case, the core-shell particles are arranged in a periodic array with sub-wavelength separations. Therefore, before moving forward, we introduce an additional analytical model accounting for the interactions among adjacent core-shell meta-atoms constituting the metasurface. This is the aim of the next sub-Section.

### *B. Metasurface homogenization*

We consider the structure shown in the inset of Fig. 3 consisting of a square lattice of core-shell spherical particles as the ones described in the previous sub-Section. We assume that the homogenization requirements are satisfied, *i.e.,* that the size of the individual particle and their separation distance are both small compared to the wavelength of the impinging field [2]. Our goal is to describe the macroscopic response of this metasurface within the frequency range between the resonance of the magnetic dipole (*md*) and the one of the electric dipole (*ed*) using effective parameters [18]-[21], assuming that the period is subwavelength and no higher-order diffraction phenomena emerge. We also assume that the array is excited by a normal-incidence plane wave so that the cross coupling between dipoles and quadrupoles is negligible.

As we have discussed in [15], an array of high-index particles can be homogenized through the coupled dipole approach and, thus, replaced by an infinitely thin sheet sustaining both electric and magnetic currents. The homogenized parameters representing the behavior of the structure are the effective symmetric $Z_{symm}$ and asymmetric $Z_{asymm}$ surface impedances, respectively. Differently from [15], however, here the individual particle has a core-shell geometry. As discussed above, this modification allows shifting the resonance frequency of the electric dipole above the one of the magnetic quadrupole, making the quadrupolar contribution non-negligible within the homogenization bandwidth. Therefore, the model developed in [15] needs to be properly generalized to account for the in-band contribution of the magnetic quadrupole.

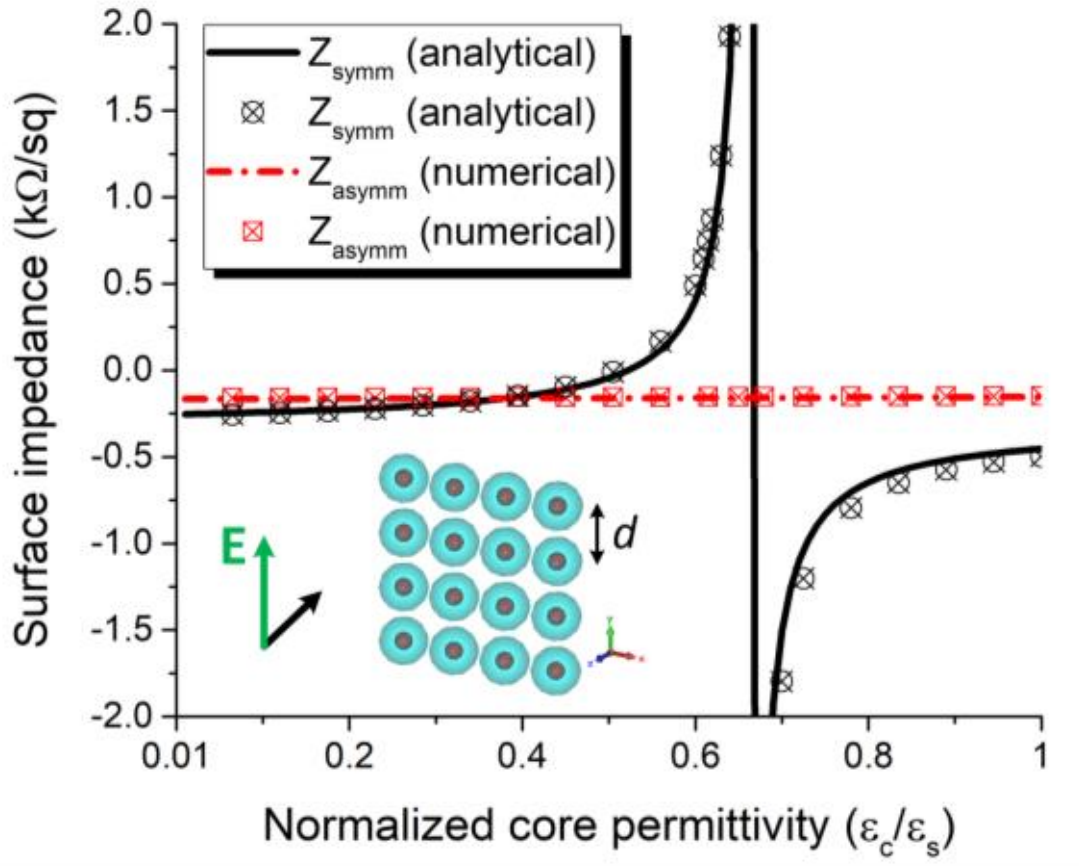

Fig. 3. Analytical (lines) and numerical (ticks) symmetric and asymmetric surface impedances as the core permittivity changes. The reference geometry is described in the main text.

According with [15], the contribution of the electric dipole to the symmetric surface impedance is equal to:

$$\frac{Z_e^d}{\eta_h} = -\frac{d^2}{k}\left[\left(\mathrm{Im}\left[\alpha_{e,d}^{-1}\right] - \frac{k^3}{6\pi}\right) + i\left(\mathrm{Re}\left[\alpha_{e,d}^{-1} - \beta_{e,d}\right]\right)\right], \quad (1)$$

where $k$ and $\eta_h$ are the wavenumber and the impedance of the host material (vacuum if we assume that the metasurface is free-standing in air or supported by a low-dielectric material), respectively, $\alpha_{e,d}$ is the dipolar electric polarizability of the core-shell sphere, and $\beta_e$ is the interaction constant among electric dipoles. By duality, we can derive a similar expression for the contribution of the magnetic dipole to the asymmetric surface admittance $Y_m^d$ depending on the dipolar magnetic polarizability $\alpha_{m,d}$ of the core-shell spherical particle and their interaction constant $\beta_{m,d}$. The expressions of the electric and magnetic interaction constant can be found in [22], whereas the dynamic polarizabilities of a core-shell particle can be related with its first two Mie scattering coefficients $a_1$ and $b_1$, *i.e.,* $\alpha_{e,d}=-i\left(6\pi/k_0^3\right)a_1$ and $\alpha_{m,d}=-i\left(6\pi/k_0^3\right)b_1$. The expressions of $a_1$ and $b_1$ for a core-shell spherical particle are available in [17].

As for the contributions to the homogenized parameters of the electric and magnetic quadrupoles, we first observe that they radiate in an opposite manner compared to the corresponding dipoles, *i.e.,* the electric (magnetic) quadrupole radiates asymmetrically (symmetrically). We also observe that the resonance of the electric quadrupole generally occurs at higher frequencies compared to both the magnetic quadrupole and the electric dipole resonances. Moreover, the core-shell geometry further shifts the electric quadrupole resonance towards higher frequencies. Since we are interested in homogenizing the metasurface in the frequency range between the resonance of the *md* and the one of the *ed*, we can safely neglect the electric quadrupole contribution and assume that the overall asymmetric surface impedance coincides with the magnetic surface impedance, *i.e.,* $Z_{asymm}=Z_m^d=1/Y_m^d$.

On the contrary, the quadrupolar magnetic resonance effectively contributes to the symmetric surface impedance of the metasurface within the homogenization bandwidth. The expression of its symmetric surface admittance $Z_m^q$ can be derived using an expression similar to (1). However, as discussed in [23], the quadrupolar magnetic polarizability $\alpha_{m,q}$ should be calculated with a slightly different expression, *i.e.,* $\alpha_{m,q}=-j\left(120\pi/k_0^5\right)b_2$. As a consequence, the multiplication coefficient outside the square bracket in (1), the imaginary part of the complex dynamic polarizability, *i.e.,* the quantity $-k^3/\left(6\pi\right)$, and the interaction constant need to be properly corrected to keep the power conservation principle satisfied [24]. The final expression of $Y_m^q$ is, thus

$$\eta_h Y_m^q=-20\frac{d^2}{k^3}\left[\left(\mathrm{Im}\left[\alpha_{m,q}^{-1}\right]-\frac{k^5}{120\pi}\right)+i\left(\mathrm{Re}\left[\alpha_{m,q}^{-1}-\beta_{m,q}\right]\right)\right], \quad (2)$$

being

$$\beta_{m,q}=\frac{jk^3}{4d^2}\left(1+\frac{1}{jk_0.R_0}\right)e^{jkR_0}. \quad (3)$$

At the end, the overall symmetric surface impedance of the all-dielectric metasurface made by core-shell meta-atoms is equal to the sum of the contributions due to the electric dipole and the magnetic quadrupole, *i.e.,* $Z_{symm}=Z_e^d+Z_m^q$.

Once the overall effective symmetric and asymmetric surface impedances have been calculated as described, it is possible to compute analytically the transmission and the reflection coefficients of the metasurface. For normal incidence, their expressions are:

$$\Gamma=\frac{-\eta_0}{2Z_{symm}+\eta_0}+\frac{Z_{asymm}}{Z_{asymm}+2\eta_0},$$
$$T=\frac{2Z_{symm}}{2Z_{symm}+\eta_0}-\frac{Z_{asymm}}{Z_{asymm}+2\eta_0}. \quad (4)$$

To check the effectiveness of the homogenization model, we have compared the analytical surface impedances in the following scenario: $r_s=0.075\lambda_0=0.75\lambda$, $r_c=(1/3)\times r_s$, $d=0.175\lambda_0$, being $\lambda_0$ the wavelength at which the results are evaluated. The analytical results have been obtained using the theory discussed above, while the numerical results have been retrieved by inverting (4) with respect to $Z_{symm}$ and $Z_{asymm}$ and using the values of $\Gamma$ and $T$ computed through numerical full-wave simulations. As it can be appreciated in Fig. 3, there is an excellent agreement between theoretical and numerical results, confirming the effectiveness of the model to analyze all-dielectric metasurfaces sustaining both dipolar and quadrupolar resonances.

In addition, as expected from the discussion available in the previous sub-Section, the homogenized parameters of the metasurface are affected in a very different way by the value of the permittivity of the spherical core. In particular, the core permittivity allows tailoring the symmetric surface impedance within a very wide range of possible values, whereas the asymmetric surface impedance remains approximately equal to the one of the metasurface composed by homogenous spherical particles ($\varepsilon_c=\varepsilon_s$). These results confirm that the core-shell geometry introduces a powerful additional degree of freedom for tailoring the reflection and transmission coefficients of dielectric metasurfaces and achieving innovative effects.

It is worth noticing that in Fig. 3 we have considered a large variation of the dielectric contrast $\varepsilon_c/\varepsilon_s$. Ultralow values of the dielectric contrast (*e.g.,* below 0.1) are useful for achieving a wide design flexibility and may be reached either by increasing the permittivity of the shell or by reducing the core permittivity. The more suitable strategy mainly depends on the considered frequencies range. In particular, at microwave and THz frequencies, where natural materials with permittivity greater than 100 are available, ultralow values of the dielectric contrast may be achieved using particles with an empty core (i.e., $\varepsilon_c$=1). Conversely, at higher frequencies (*e.g.,* near-IR and visible spectrum), where the refractive index of natural materials is limited [25], the design of a core made by a low-losses epsilon-near-zero metamaterial would be required [26]-[28]. In the following, we consider the former scenario and focus our attention on the design of dielectric metasurfaces made by spherical particles with an empty core and a high-index shell.

## III. Design of Pass-Band Filters

In this Section, we describe how it is possible to obtain an ultra-narrow passband behavior through an all-dielectric metasurface by exploiting the resonances of the electric and magnetic dipole and the one of the magnetic quadrupole.

The first step consists in designing an all-dielectric metasurface made by spherical *homogeneous* particles and behaving as an ultra-thin dielectric reflector. This possibility has been first discussed in [8], whereas a careful design approach based on surface impedance homogenization is reported in [15]. In short, the idea is to exploit the strong interactions between electric and magnetic dipoles sustained by the spherical particles to approach the electric and magnetic resonances and achieving a large reflection bandwidth rather than two isolated reflection peaks. The typical reflection coefficient obtained by this structure is shown in Fig. 4 (dashed red line) and exhibits relevant reflection performances (R > -1 *dB*) within the range $0.68f_0$-$1.02f_0$, corresponding to a fractional bandwidth of 40%. Compared to this result, the new possibilities enabled by core-shell particles are: *i)* widening the reflection bandwidth exploiting three modes rather than two; *ii)* opening a narrow transmission peak within the reflection bandwidth by exploiting the resonance of the magnetic quadrupole.

For these purposes, we have designed an all-dielectric metasurface made by core-shell meta-atoms with the following parameters: $r_s = 0.069\lambda_0 = 0.69\lambda$, $r_c = (0.39)\times r_s$, $d = 0.138\lambda_0$. To achieve these values, we have first fixed the contrast between the permittivity of core and shell (*e.g.,* $\varepsilon_s = 100$ and $\varepsilon_c = 0.1\varepsilon_s$) and then carefully adjusted the size of the meta-atoms and their periodicity to obtain two adjacent wide reflection bands (Fig. 4, in solid black line). The first one (from $0.68f_0$ to $0.989f_0$) is due to the interaction between the magnetic dipole and the magnetic quadrupole and almost overlaps with the reflection band achieved in the homogenous case; the second one (from $1.01f_0$ to $1.26f_0$) is a new reflection band caused by the interaction between the magnetic quadrupole and the electric dipole, which is now shifted toward higher frequencies. As expected, two important differences compared to the homogenous case can be observed: first, the -1*dB* reflector bandwidth is wider (59% vs. 40%) because three different modes are exploited to achieve full reflection rather than two. More importantly, the presence of the magnetic quadrupole resonance in between the resonances of the electric and the magnetic dipole allows achieving a balanced condition between symmetric and asymmetric scatterers [23] and opening a narrow full-transmission window in the middle of the reflection bandwidth. The -3 *dB* fractional bandwidth is narrow and equal to 2.1%, corresponding to a remarkable Q-factor of 48. The Q-factor can be further increased by using a shell dielectric with higher permittivity: for instance, it reaches the value of 80 when the dielectric shell is made by a material with permittivity $\varepsilon_s = 150$.

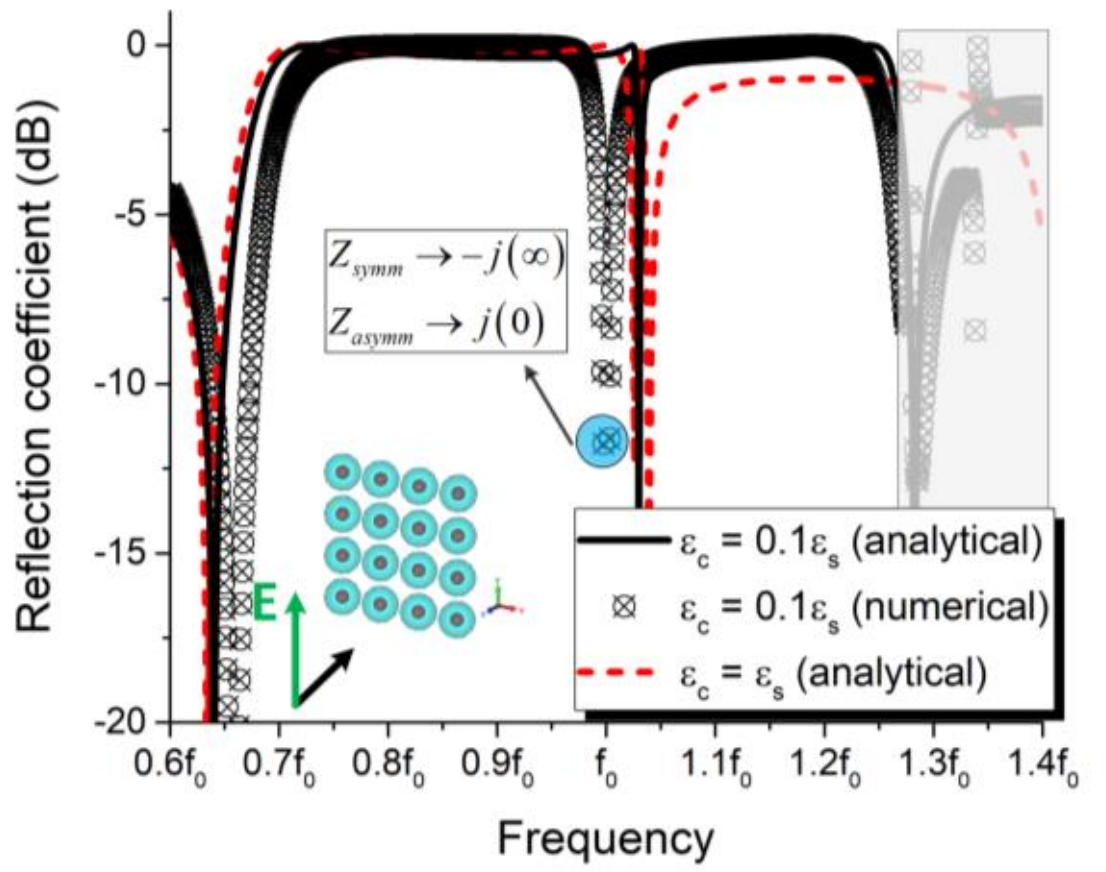

Fig. 4. Reflection coefficient of an all-dielectric metasurface composed by core-shell particles (solid line) *vs.* the one achieved using an equivalent all-dielectric metasurface composed by bulk spheres (dashed line). Ticks represent the results of full-wave simulations. The grey area of the plot represents the frequency region beyond the electric dipole resonance in which the homogenization requirements are not satisfied due to the emergence of higher-order resonances.

As it can be appreciated in Fig. 4, there is a good agreement between numerical and analytical results. The only difference is a slight frequency shift of the magnetic resonances (of both the dipole and the quadrupole). This discrepancy, which is lower than 3%, is mainly due to the very strong interactions between the particles that, in the present case, are touching each other (*i.e.,* $d=2\times r_s$). This scenario does not strictly satisfy the homogenization requirements of the coupled-dipole approach [18]. Nevertheless, the obtained results are still very effective and allows achieving an initial design that can be straightforwardly optimized through a quick run of full-wave simulations.

So far, we have not discussed in detail the possible materials that can be used to achieve the aforementioned effects. The material's choice strongly depends on the frequency range of interest. As a rule-of-thumb, to efficiently excite both the electric and the magnetic resonances, a spherical particle should be made by a material with a refractive index higher than 4 [8]. These values can be easily obtained, even at optical frequencies, using semiconductors such as Ga-P or c-Si [25]. However, as we have demonstrated with the previous example, ultralow values of the dielectric contrasts between the core and the shell are often required. At microwave and THz frequencies, ceramic materials [30], extensively developed since the '80, allow achieving $\varepsilon_c / \varepsilon_s < 0.1$ even by considering particles made by an empty core. For example, some commercially available products made by magnesium calcium titanate exhibit a real part of the permittivity ranging from 18 to 140 with dielectric loss tangent lower than 0.0015. Similar values can be achieved even at THz frequencies using, for instance, zirconium tin titanate [30]. These ceramic materials can be realized using solid-state reaction technique, which results into a fine powder to be pressed into a desired shape with high accuracy [31].

An interesting related analysis can be carried out about the robustness of the devices toward potential manufacturing inaccuracies. As discussed in our earlier works on plasmonic metasurfaces [19]-[21] and in literature [32]-[33], metasurface are generally quite robust towards fabrication imperfections because their response is averaged on all the composing meta-atoms. We have verified this feature also for the design under consideration and the results (not shown here for the sake of brevity) confirm that the operation frequency of the filter is almost insensitive to realistic geometrical deformations and material perturbations, even in conservative scenarios (*i.e.*, more than 25 % of the particles exhibit either a geometrical or a material perturbation). Notably, the material perturbation plays a more important role, as moderate values (around 1.5% variation of $\varepsilon_s$ in 8% of the particles) introduce ultra-narrow spikes in the reflection coefficient. To this regard, however, we emphasize that the homogeneity level of the electromagnetic properties of high-index ceramics is extremely high and may be below 0.13‰ [30].

Before concluding this Section, we investigate the performances of the designed filter *vs.* the losses of the dielectric material and the impinging angle. We consider the metasurface shown in the inset of Fig. 4 designed to operate at microwave frequencies. The shell is assumed to be made by a ceramic material with permittivity $\varepsilon_s = 100$ and loss tangent *tan(δ)*, whereas the core of the particle is made by vacuum, *i.e.,* $\varepsilon_c = 1 = 0.1\times\varepsilon_s$. In Fig. 5, we report the magnitude of the transmission coefficient obtained through full-wave simulations for different values of *tan(δ) vs.* the impinging angle $\theta$. The solid lines refer to the TM polarization, whereas the dashed lines to the TE polarization. As it can be appreciated, despite the geometrical symmetry of the particles, the filter exhibits a different behavior for the two polarizations for off-normal incidence. In particular, a narrow angular bandwidth ($T$ >-3 *dB* for $\theta \in$ [-30°, 30°]) can be observed for the TE polarization. This effect is due to the progressive frequency shift of the magnetic quadrupolar resonance induced by the variation of the orientation of the impinging magnetic field. Conversely, the TM polarization is almost unaffected by the impinging angle due to the absence of the magnetic field component in the longitudinal direction. By duality, the TE polarization would not be affected by a variation of the angle $\varphi$. It is worth mentioning that the strong spatial dispersion of the proposed all-dielectric filters may be exploited for designing space polarizers working for extreme angles of incidence.

From Fig. 5, in addition, we may also appreciate the robustness of the filter towards the EM losses of the dielectric material composing the particle shell. The transmission for normal incidence, in fact, is above 0.9 for a *tan(δ)* lower than 0.001. These values are compliant with the ones of ceramic materials at microwave and infrared frequencies [30], making these structures realizable with real materials.

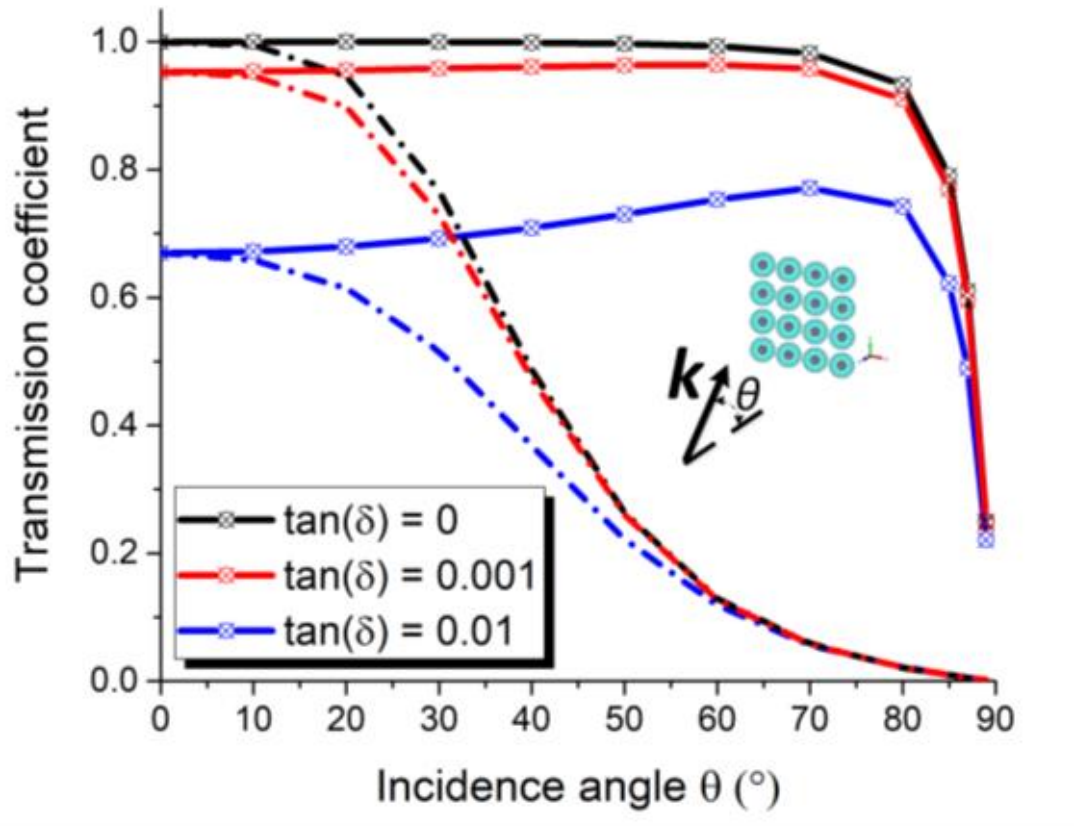

Fig. 5. Angular behavior of the transmission coefficient of the all-dielectric metasurface within its passband for both TM (solid lines with ticks) and TE (dot-dashed lines) polarization. The results are shown for different values of the loss tangent of the dielectric shell.

## IV. Self-Filtering Horn Antennas

In this Section, we explore one of the possible applications of the proposed filters. In particular, we focus our attention to the design of self-filtering antennas (also known as filtennas), *i.e.,* EM components integrating both a filtering and a radiating module. These structures exhibit a narrower impedance bandwidth compared to conventional antennas and, therefore, are able to reject the out-of-band noise and potential interfering signals. In the past years, several horn filtennas have been proposed [34]-[37], especially for receiving satellite systems, but the filtering behavior has been achieved relying on the use of resonant metallic inclusions. Here, we show how an all-dielectric filter allows designing self-filtering horn antenna with enhanced out-of-band rejection capabilities.

The structure we consider is shown in the inset of Fig. 6: it consists of a regular pyramidal horn (140×103 $mm^2$) antenna excited by a WR90 waveguide (22.86×10.16 $mm^2$) and working in the X-band (*i.e.,* magnitude of the reflection coefficient lower than -20 *dB* within the frequency range [8.2-12.4] *GHz*). In order to reduce its operative bandwidth within a narrow frequency range around the central frequency, we have loaded the throat of the horn with a single core-shell particle. As can be appreciated in the inset of Fig. 6, the particle is placed inside the hole of a metallic wall. This situation emulates the periodic metasurface we have investigated in the previous Sections and allows using the analytical model developed above. In addition, in this scenario, the filtering module can be considered as an add-on element to be plugged/removed to/from the horn aperture, depending on the needs. However, we underline that similar filtering effects can be also achieved by removing the metal wall and replicating the designed core-shell particle in a regular lattice designed in such a way to fit the feeding aperture of the horn antenna.

The final design is characterized by the following parameters: $r_s$ = $0.068\lambda_0$ and $r_c = 0.39\times r_s$, being $\lambda_0$ the wavelength at the central frequency of the X-band. The material used for realizing the external shell is a ceramic with relative permittivity $\varepsilon_s = 100 - j(0.1)$. The magnitude of the reflection coefficient at the antenna input horn is shown in Fig. 6. As it can be appreciated, the impedance bandwidth of the filtenna is greatly reduced compared to the unloaded horn. In particular, its fractional bandwidth is 1.4%, corresponding to an absolute bandwidth of 140 MHz, significantly more selective than the one achieved by previous designs discussed in the literature. The out-of-band rejection capabilities of the proposed filtenna are confirmed by the analysis of its broadside directivity and realized gain (*i.e.,* the product between the gain and the matching losses), both reported in Fig. 7 as the frequency changes.

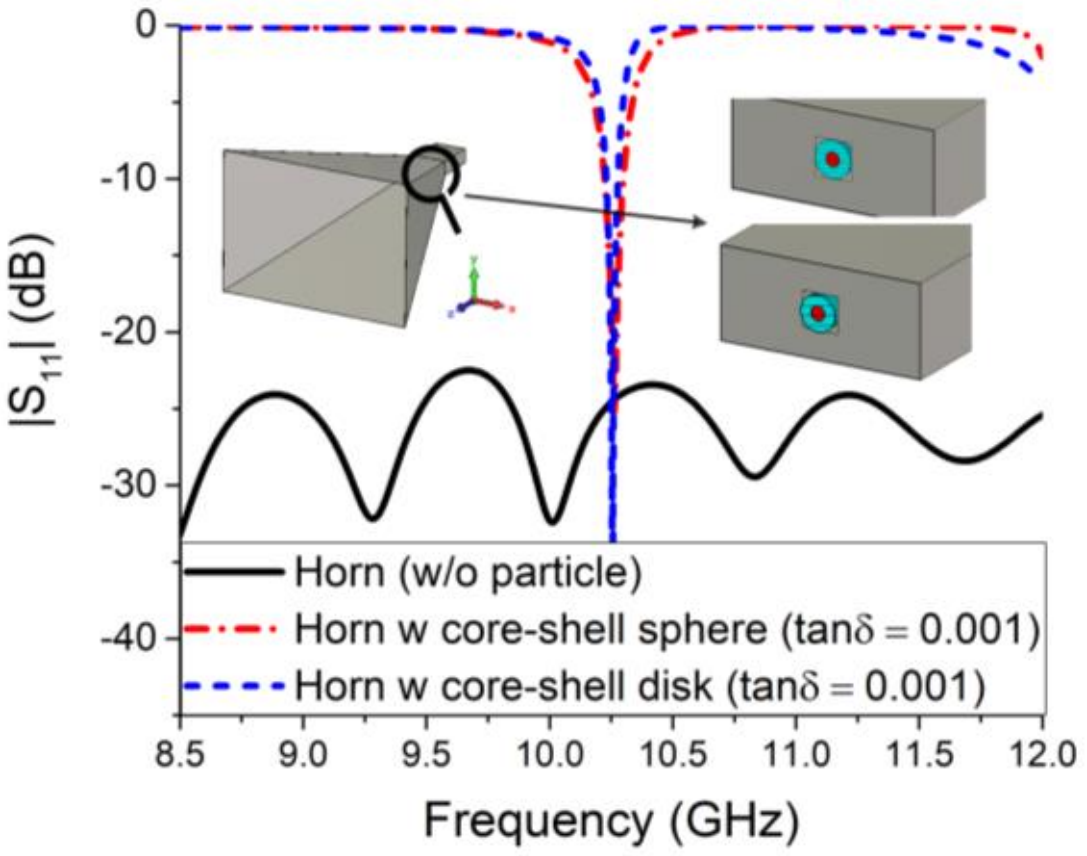

Fig. 6. Magnitude of the reflection coefficient of the horn antenna loaded with two different core-shell particles compared to the one of the unloaded antenna.

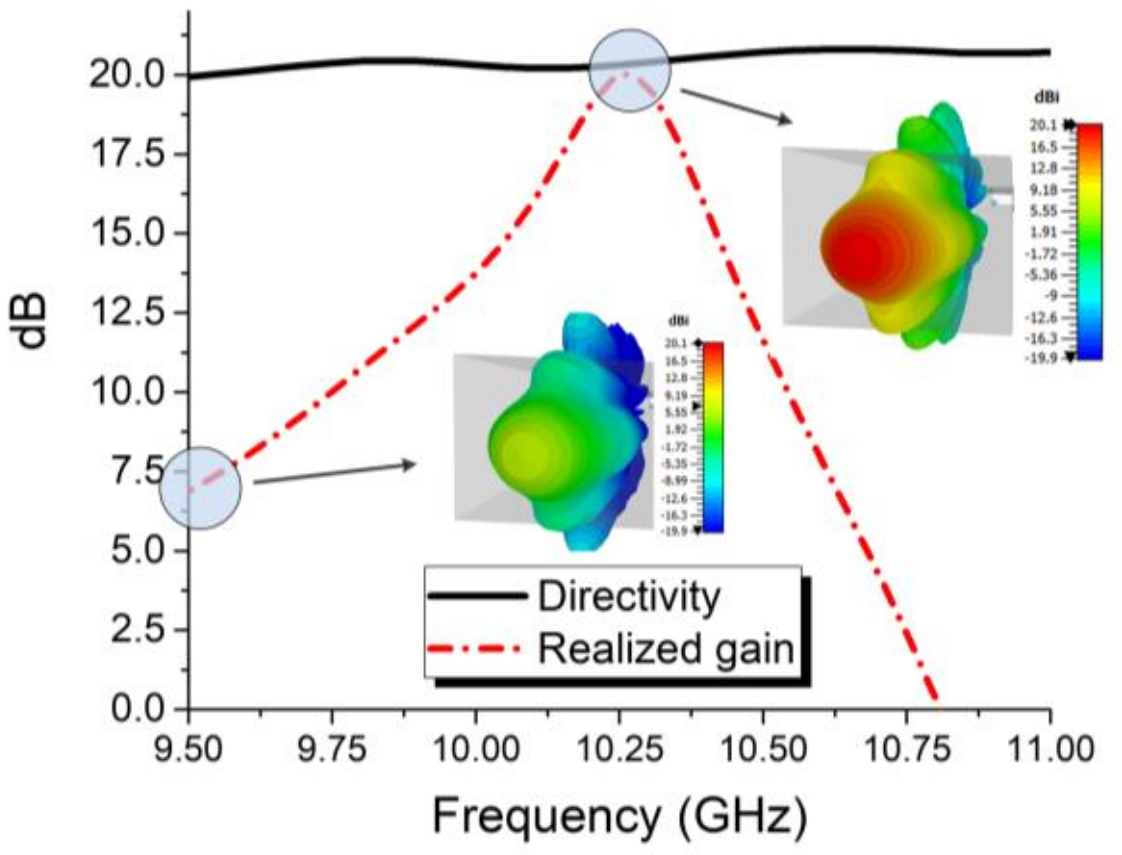

Fig. 7. Broadside directivity and realized gain of the proposed self-filtering horn antenna as a function of frequency. In the insets, the three-dimensional realized gain patterns at two different frequencies are reported.

As it can be appreciated, the directivity of the self-filtering horn filtenna is almost unaffected by the presence of the core-shell particle (D ≈ 20 dB within the entire X-band, as the original horn). On the contrary, because of the filtering effects of the core-shell particle on the antenna reflection coefficient, the realized gain of the self-filtering horn shows a high frequency selectivity, *i.e.,* a 3*dB* fractional bandwidth equal to 2.3%.

Finally, in view of a possible experimental realization, we have investigated the filtering performances of a slightly different filter consisting of a core-shell disk. As well known, the use of disks may relax the fabrication constraints because of their open shape. Even though disks exhibit a more complex scattering response than spheres due to the presence of additional Fabry-Perot resonances [38], they can still be optimized to achieve the required balance between dipoles and quadrupoles. In this case, however, the final design can be achieved only numerically because closed-form expressions of the disk polarizabilities are not available. As we have proven earlier [19], the analytical design obtained for spheres can be used as an effective starting point of the disk optimization process. The full-wave results for a core-shell disk with size $r_s = 0.066\lambda_0$, $r_c = 0.4\times r_s$ and $L = 2.28\times r_s$ are shown in Fig. 6. As it can be appreciated, excellent filtering performances are achieved, confirming the possibility to replace spheres with disks in practical scenarios.

## V. Conclusion

In this work, we have described a novel approach to design pass-band space filters using all-dielectric metasurfaces. The approach relies on the proper balance of both electric and magnetic resonances sustained by a lattice of dielectric resonators. Specifically, it has been shown that, by using a core-shell geometry as individual meta-atom, it is possible tailoring the resonances of the electric dipole, magnetic dipole, and magnetic quadrupole to design a wideband mirror with a narrow transmission passband. The robustness of the design versus the impinging angle and the dielectric losses has been discussed. In addition, for design purposes, a reliable homogenization model that accounts for the effect of the magnetic quadrupole has been developed and numerically verified. Finally, a possible application of the proposed filter for the design of self-filtering horn antennas has been discussed and numerically verified.